\documentclass{sig-alternate} 
\usepackage{booktabs}
\usepackage{datetime}
\usepackage[ruled,vlined,linesnumberedhidden,lined,boxed,commentsnumbered]{algorithm2e}
\usepackage{hyperref}
\usepackage{multirow}
\usepackage{dsfont}
\usepackage{array}
\usepackage{pifont}
\usepackage{tabularx}
\usepackage{enumitem}
\usepackage{makecell}
\usepackage{cellspace}
\usepackage[T1]{fontenc}

\usepackage{color, colortbl}

\newfont{\mycrnotice}{ptmr8t at 7pt}
\newfont{\myconfname}{ptmri8t at 7pt}
\let\crnotice\mycrnotice%
\let\confname\myconfname%
 \DeclareMathOperator{\1}{\mathds{1}}

\permission{Permission to make digital or hard copies of all or part of this work for personal or classroom use is granted without fee provided that copies are not made or distributed for profit or commercial advantage and that copies bear this notice and the full citation on the first page. Copyrights for components of this work owned by others than ACM must be honored. Abstracting with credit is permitted. To copy otherwise, or republish, to post on servers or to redistribute to lists, requires prior specific permission and/or a fee. Request permissions from Permissions@acm.org.}

\CopyrightYear{2016}
\setcopyright{acmcopyright}
\conferenceinfo{CIKM'16 ,}{October 24-28, 2016, Indianapolis, IN, USA} \isbn{978-1-4503-4073-1/16/10}\acmPrice{\$15.00} \doi{http://dx.doi.org/10.1145/2983323.2983808}

\begin{document}
\makeatletter
\newcommand\footnoteref[1]{\protected@xdef\@thefnmark{\ref{#1}}\@footnotemark}
\makeatother
\newcommand{\seq}[1]{\langle\,{#1}\,\rangle}
\newcommand{\set}[1]{\left\{\,{#1}\,\right\}}
\newcommand{\bigset}[1]{\left\{\,{#1}\,\right\}}
\newcommand{\bigtuple}[1]{\left(\,{#1}\,\right)}
\newcommand{\card}[1]{\left|\,{#1}\,\right|}
\newcommand{\tup}[1]{\left(\,{#1}\,\right)}
\newcommand{\argmin}[2]{\underset{#1}{\operatorname{arg\,min}}{\:\: #2}}
\newcommand{\stdmin}[2]{\underset{#1}{\operatorname{min}}{\:\: #2}}
\newcommand{\simplemin}[1]{\ensuremath{\underset{#1}{min}\;}} 
\newcommand{\pow}[2]{\ensuremath{#1^#2\;}}
\newcommand{\ra}[1]{\renewcommand{\arraystretch}{#1}}
\newcommand {\N}{\mathcal{N}}
\newcommand{\argmax}{\operatornamewithlimits{argmax}}
\newcolumntype{P}[1]{>{\centering\arraybackslash}p{#1}}

\hyphenation{Map-Reduce}  
\hyphenation{opti-mi-za-tion}
\hyphenation{Wiki-pedia}
\hyphenation{sali-ence}
\hyphenation{Guard-ian}
\hyphenation{launch-ed}
\hyphenation{foot-ball}

\title{Finding News Citations for Wikipedia}
\numberofauthors{1} \author{ \alignauthor
  Besnik Fetahu$^\dagger$, Katja Markert$^\ddagger$, Wolfgang Nejdl$^\dagger$, Avishek Anand$^\dagger$\\
  \affaddr{
    \begin{tabular}{c@{~~~~}c}
      \@{}&\@{}\\
      $^\dagger$L3S Research Center &  $^\ddagger$Institute of Computational Linguistics\\
      Leibniz University of Hannover & Heidelberg University\\
      Hannover, Germany & Heidelberg, Germany\\
      \affaddr{\{fetahu,nejdl,anand\}@L3S.de} & \affaddr{markert@cl.uni-heidelberg.de}
    \end{tabular}
  }
}

\maketitle

\begin{abstract} 

An important editing policy in Wikipedia is to provide citations for added statements in Wikipedia pages, where statements can be arbitrary pieces of text, ranging from a sentence to a paragraph. In many cases citations are either outdated or missing altogether. 

In this work we address the problem of finding and updating news citations for statements in entity pages. We propose a two-stage supervised approach for this problem. In the first step, we construct a classifier to find out whether statements need a news citation or other kinds of citations (web, book, journal, etc.). In the second step, we develop a news citation algorithm for Wikipedia statements, which recommends appropriate citations from a given news collection. Apart from IR techniques that use the statement to query the news collection, we also formalize three properties of an appropriate citation, namely: (i) the citation should entail the Wikipedia statement, (ii) the statement should be central to the citation, and (iii) the citation should be from an authoritative source. 

We perform an extensive evaluation of both steps, using 20 million articles from a real-world news collection. Our results are quite promising, and show that we can perform this task with high precision and at scale.
\end{abstract}

\section{Introduction}\label{sec:introduction}

Wikipedia has become the most used Internet encyclopedia and, indeed,
one of the most popular websites overall.\footnote{\small{In 2015 it
    was in the top 10 most visited Internet sites according to the
    Alexis Internet ranking \url{www.alexa.com}).}} In addition, due
to Wikipedia's inclusion into widely used applications such as Google
KnowledgeGraph or Apple's Siri system, its content will influence the
knowledge and, potentially, the behavior of millions of users, even if
they do not visit the Wikipedia site directly. Therefore, it is
essential that its content is accurate and reliable.

In contrast to traditional encyclopedias,  Wikipedia is not authored mainly by
experts. Also, the articles are authored collaboratively by more than just a
small number of contributors and the identity and expertise of authors
is hard to verify. This leaves Wikipedia articles open to
addition of inaccurate content, spamming or vandalism, and
calls into question its reliability. A substantial number of
reliability studies have compared Wikipedia against other reference
works (such as the \textit{Encyclopedia Britannica} or drug package
information) or subjected them to expert review: The exhaustive survey
in \cite{ASI:ASI23172} concludes that the results of these studies
have overall been favourable to Wikipedia when it comes to accuracy of
facts, although some works (especially on medical articles) found
errors of omission.\footnote{\small{The standard for medical
    information should be higher for obvious reasons and omitted
    information for side effects or risks can be crucial.}}

These surprisingly favorable results on the reliability of Wikipedia
can in all probability be traced to a small number of Wikipedia
editorial policies, one of which we are concerned with in this
paper. The \textit{Verifiability} policy requires Wikipedia
contributors to support  their additions with citations from
authoritative external sources. In particular, Wikipedia policy states
that ``articles should be based on reliable, third-party, published
sources with a reputation for fact-checking and
accuracy.''\footnote{\label{reliability}\small{\url{https://en.wikipedia.org/wiki/Wikipedia:Identifying_reliable_sources}}}
This policy, on the one hand, guides contributors towards both
neutrality and the importance of authoritative assessment and, on the
other hand, allows Wikipedia core editors to identify unreliable
articles more easily via a lack of such citations. Citations therefore
play a crucial role in ensuring and upholding Wikipedia reliability.

For current and recent events, news citations are one of the most-used sources \cite{fetahu2015much}.  Again, Wikipedia encourages the use of news outlets as citations: ``news reporting from well-established news outlets is generally considered to be reliable for statements of fact''\footnoteref{reliability}. As we show in Section~\ref{sec:datasets_groundtruth}, news are indeed the second-most widely used citation category in Wikipedia (with 1.88 million citations in our English Wikipedia snapshot) -- however, around 26\% of these are no longer available due to dead or redirected links.  In addition, new information is added all the time and will need verification. For both these purposes, an automatic way of finding an authoritative news citation for any fact(s) one might wish to update, locate again or add would greatly facilitate Wikipedia editing and improve its reliability. Moreover, if no such citation can be found, it can guide contributors or core editors towards questioning their edits.

In this paper, we suggest such a method for automatic news citation
discovery for Wikipedia. In particular, we make the following
contributions: (i) We analyze for which type of Wikipedia statements a
news citation is appropriate (in contrast to, for example, a
scientific journal citation), taking into account the type and
structure of entity the statement is about, as well as the language
the statement is written in. We provide a supervised learning
algorithm for  statement classification into citation categories.
(ii) We then develop a citation discovery algorithm which formalizes
three properties of a good citation, namely that it entails the
statement it supports, that it is from an authoritative source and
that the statement it supports is central to it. (iii) We establish a
large-scale evaluation framework for citation discovery which uses
crowdsourcing for measuring our approach's precision.

To the best of our knowledge, this is the first work that
automatically discovers citations for fine-grained Wikipedia
statements. We show that news citations can be discovered with high
precision, in large contemporary news collections. In particular, we
with high accuracy recover the same or very similar citations as the
ones originally given by Wikipedia contributors in the presence of
numerous strong distractors or even find citations which are
preferable to the original ones (as established via crowdsourcing).

\section{Problem Definition and\\ Approach Outline}\label{sec:problemdefinition}

In this section, we describe the terminology and problem definition
for finding \emph{news citations} for Wikipedia.

\subsection{Terminology and Problem Definition}\label{subsec:problem_definition}

We operate on a specific snapshot of Wikipedia $\mathbf{W}$ where the
text in each Wikipedia page $e \in \mathbf{W}$ is organized into
\emph{sections} denoted by $\Psi(e)$. Additionally, entity pages are
organized into a \emph{type} structure, which is a
\emph{directed-acyclic-graph} (DAG) induced by the Wikipedia
categories. This is routinely exploited by knowledge bases like YAGO
(e.g. \emph{Barack\_Obama} \texttt{isA Person}) and we leverage this
type structure where each page $e$ belongs to a set of types
$T(e)$. We, however, modify the original YAGO type structure to make it
\emph{depth consistent} as explained in
Section~\ref{subsec:learning_framework}.

\subsubsection{Citations and Wikipedia Statements}
\label{sec:statements_citations}

\textbf{Citation:} In Wikipedia pages, any \emph{piece of text} can be
supported by a \emph{citation}. The citation points to an external
information source, such as a news article, blog, book or journal,
that is considered as \emph{evidence} for the fact mentioned in the
text. Citations in Wikipedia are categorized into a predefined set of
\emph{16 citation categories} viz. $c=\{$\texttt{web}, \texttt{news},
\texttt{books}, \texttt{journal}, \texttt{map}, \texttt{comic},
\texttt{court}, \texttt{press release}, $\ldots\}$. The distribution
of the citation types is given in Figure~\ref{fig:cite_type_dist}.

\textbf{Statement:} We will refer to the \emph{piece of text} from a
Wikipedia page that has or needs a citation as a \emph{Wikipedia
  statement} or simply a \emph{statement}. In this work, we restrict
statements to a single sentence or a sequence of
sentences that occur between two consecutive citation markers or a
citation marker and paragraph beginning/end.  A citation marker is
either an actual citation or a placeholder \texttt{citation
  needed}\footnote{\small{\url{https://en.wikipedia.org/wiki/Template:Citation_needed}}}. We
therefore leave the identification of statements to future work.  We
also do not consider finding evidence for partial sentences or
clauses.  Each statement $s$ in a page $e$ belongs to a section $\psi
\in \Psi (e)$, and the set of statements extracted from a section
$\psi$ of $e$ is represented as $S(e,\psi)$.

\textbf{Anchors and Entities: } Typically words or phrases in statements
link to other Wikipedia pages which represent entities through
\emph{anchors}. We denote these links to other pages or entities
starting from a statement $s$ as $\gamma(s)$, and $T(s) \,\,=\,\, \{
T(e) \,|\, e \in \gamma(s) \}$ the corresponding entity types.

\subsubsection{Citation Finding Tasks}\label{subsubsec:citation_finding_tasks}

We posit that the following two tasks are integral to finding a
citation for a Wikipedia statement.

\textbf{Statement Categorization.} For a statement $s$ from a page $e$
of an unknown citation category, the task aims to determine the
correct citation category for $s$.
\begin{equation}
SC:  f(s,e) \rightarrow c, \text{where $c\in\{$\texttt{web, news,$\ldots$}}\}
\end{equation}

We want to categorize $s$ as a \emph{news statement} if it requires a news citation. This is based on the hypothesis that each statement typically has a preferred citation category, which we need to determine before making a high precision citation recommendation.

\textbf{Citation Discovery.} Given a (i) statement $s$ found in page
$e$ and of category $c=$\texttt{news}, and (ii) an external news
collection $\N$, we define the citation discovery task as finding
articles $n \in \N$ that serve as evidence for $s$. We define the
function $FC$ which for $s$ outputs the subset of articles that can be
suggested for citation.
\begin{equation}
FC: f(s,e,\mathcal{N}) \rightarrow \langle s, n\rangle \in \{`correct', `incorrect'\}
\end{equation}

\subsection{Approach Overview}\label{subsec:approach_overview}

\begin{figure*}[ht!]
  \centering
  \includegraphics[width=0.8\textwidth]{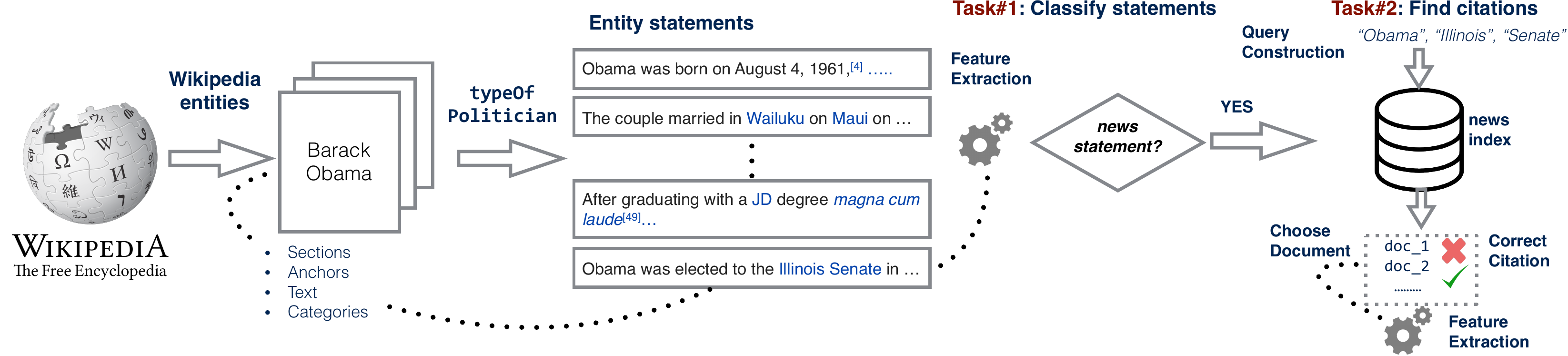}
  \caption{\small{Approach overview. In the first task, we classify statements into one of the citation types, while in the second we find the missing citation for statements of type \emph{news}.}}
  \label{fig:approach_overview}
\end{figure*}

Figure~\ref{fig:approach_overview} shows an overview of our approach. For an entity, we extract entity and type structure, and its statements and finally run the steps of statement categorization and citation discovery.

\textbf{\emph{Statement Categorization--SC.}} In the first step, we
predict the citation category of a Wikipedia statement $s$ via
supervised machine learning. We train a multi-class classification
model, where the classes correspond to the citation categories $c$.

\textbf{\emph{Citation Discovery--FC.}} In the second step, for all
\emph{news statements} we find evidence for them via news articles. We
retrieve candidate news articles from a news collection
$\N$ through standard \emph{information retrieval} methods with $s$
serving as our query, and classify each candidate as either an
appropriate citation for $s$ or not.

\section{Wikipedia Ground-Truth}\label{sec:datasets_groundtruth}

\subsection{Ground-Truth: Wikpedia News Statements}
From a Wikipedia snapshot $\mathbf{W}$ (2015-07-01) we extract all
statements and all citations associated with that
statement.\footnote{As a statement can have different clauses,
  sometimes extracted citations only serve as evidence for part of the
  statement.  We, however, do not distinguish at this level of
  granularity but assume that all associated citations support the
  whole statement.}  We extract 6.9
million statements with 8.8 million citations, from 1.65
million entities and 668k section types.

Citations are categorized into one of the
categories $c$ by the Wikipedia editors.
However, sometimes the editors do not categorize a citation as \texttt{news}
although they should do so. 

For example, in \textbf{W}, its top--3 news domains \emph{BBC}, \emph{NYTimes}, \emph{Guardian}, are often cited in categories other than \texttt{news}.\footnote{Thus, the citation \small{\url{http://news.bbc.co.uk/1/hi/uk_politics/7433479.s\ tm}}  from the entity \texttt{Liam} \texttt{Byrne} has been categorized as \texttt{web}, although the more specific \texttt{news} category  would have been appropriate.}
Most of such \emph{violations} by the editors occur when citing news under the category \texttt{web}, which often is a catch-all for almost any type of resource (news, book, etc.).

In most cases such violations can be accurately corrected by applying
two simple heuristics:

\textbf{Majority Voting.} Citations from the same domain URL are
tagged with different categories. We resolve such cases based on
\emph{majority voting}. In case a domain is cited more often under the
\texttt{news} category, then all citations to the same domain are
changed to \texttt{news}.

\textbf{URL Patterns.} In this heuristic we look for patterns in the
URL, specifically for `\emph{/news/}' and `\emph{http://news.}'. This
rule is applied to \texttt{web} statements, and in case the URL
matches one of the patterns, we change its category to \texttt{news}.

Table~\ref{tbl:gt_cite_changes} shows the top--4 most frequent citation categories and the impact of our ground-truth curation rules. Rule application changes the citation category for 1,652,619 citations, approximately 18\% of all citations in \textbf{W}. The cells in the table show the number of statements that are changed from the category in the row to the category in the column table.

\begin{table}[ht!]\small
\centering\scalebox{0.9}{
\begin{tabular}{l l l l l }
\toprule
& book & journal & news & web\\
\cmidrule{2-5}
book & 0 & 2,650 & 1,155 & 71,801\\
journal & 14,905 & 0 & 13542 & 110,133\\
news & 5,698 & 2,770 & 0 & \textbf{391,634}\\
web & 16,549 & 25,109 & \textbf{944,977} & 0\\
\bottomrule
\end{tabular}}
\caption{\small{The cells show the number of statements that are changed from one category to another category after ground-truth curation.}}
\label{tbl:gt_cite_changes}
\end{table}

\begin{figure}[ht!]
\centering
\includegraphics[width=0.8\columnwidth]{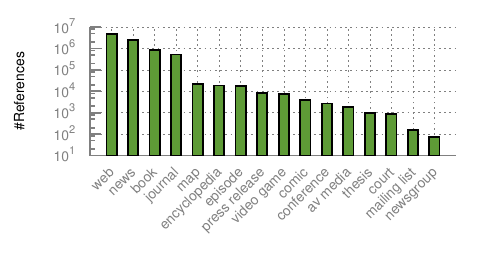}
\caption{\small{Statement distribution by citation category.}}
\label{fig:cite_type_dist}
\end{figure}

We say that a statement is a \emph{news statement} if it contains at
least one \texttt{news} citation (after ground-truth
curation). Figure~\ref{fig:cite_type_dist} shows the statement
distribution across the categories. It is evident that \texttt{web}
and \texttt{news} are the two most popular categories, with 5.3 and
1.88 million citations, coming from  1.2 million and 436k
entities, respectively.

\subsection{Wikipedia News Collection}\label{subsec:data_wiki_news}

From the \emph{news statements}, we extract the cited news articles
and construct the Wikipedia news collection $\mathcal{N}^{W}$, which
serves as our ground-truth for the \emph{citation discovery} task. We
define $N_t\subseteq\N^W$ as the set of articles cited from statements
$s$ which come from entities of type $t$. With $N_s$ we denote the set
of articles cited by $s$.

\begin{figure}[ht!]
\centering
\includegraphics[width=0.8\columnwidth]{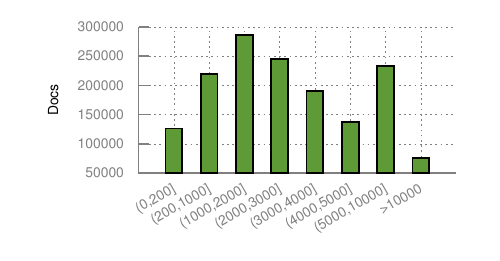}
\caption{\small{News article length (in number of characters) distribution.}}
\label{fig:news_length}
\end{figure}

From the collection of news statements, we have 1.88 million citations
to news articles (see above). We successfully crawled 1.5 million
articles.The remaining  19\% of citations point to non-existent
articles (dead links, moved content etc.). Furthermore, some of the
successfully crawled URLs point to
the \emph{index pages}. This can be noticed  when we
consider the article length (in terms of characters) in
Figure~\ref{fig:news_length}. Filtering out articles that are
below 200 characters, we are left with with 1.39 million articles, a
decrease of 26\% from the original 1.88 million news citations.

An additional issue we notice in $\N^W$ are citations to non-English news articles. We find that 23\% of articles in $\mathcal{N}^W$ are in languages other than English, using Apache Tika\footnote{\small{\url{http://tika.apache.org}}} for language detection.

\section{Statement Categorization}\label{sec:statement_classification}

In the statement categorization task, we are given a statement $s$ and
the entity $e$ from which it is extracted. We compute features that
exploit the language style of $s$ and the type and section structure
of $e$ to categorize $s$ into one of the citation categories $c$. We
learn a multi-class classifier
(Section~\ref{subsec:learning_framework}) with classes corresponding
to citation categories $c$ and optimize for predicting news
statements.  Table~\ref{tbl:sc_feature_list} shows an overview of the
feature list.

\begin{table}[ht!]\small
\centering
\scalebox{0.9}{
\begin{tabular}{p{2.3cm} p{4.5cm} p{1.5cm}}
\toprule
feature & description &\\
\midrule
$\# verbs\_attr$ & the number of verbs of attribution & \multirow{5}{1.5cm}{\emph{Language Style}} \\
$\# POS$ & the frequency of POS tags in $s$ & \\
$\lambda(s)$ & temporal proximity of $s$ to time point &\\
$discourse$ & discourse annotations of $s$ & \\
$\#quotations$ & the frequency of quotations in $s$ & \\
$\theta(s,N_t)$ & LM score of $s$ & \\
$LDA(s, N_t)$ & similarity of $s$ to a topic model & \\[1.5ex]

$p(s= news | \psi)$ & \multirow{2}{4.5cm}{section and type news-priors, with \emph{min, max} and \emph{avg} scores of $p(s=news | \psi)$ and $p(s=news | t)$ for $e$ }&  \multirow{4}{1.5cm}{\emph{Entity Structure}}\\
$p(s=news | t)$ & & \\[2ex]
$p(s=news | t',t)$ & type co-occurrence probability between $t\in T(e)$ and $t'\in T(s)$ & \\[1.5ex]
$p(s=news|t,\psi)$ & type-section joint probability scores & \\
\bottomrule
\end{tabular}}
\caption{\small{Feature list for statement categorization.}}
\label{tbl:sc_feature_list}
\end{table}

\subsection{Statement Language-Style}

We hypothesize that Wikipedia statements with news citations are
similar to the language style of news, as they often paraphrase cited
news articles.  Different genres (such as \textit{news, recipes,
  sermons, FAQs, fiction \ldots}) differ in their linguistic
properties as the different functions they fulfill influence
linguistic form \cite{biber1991variation}. For example, we expect news reports (which center mostly on past events) to contain
more past tense verbs than a recipe which gives instructions via verbs
in the imperative. We use features that were successful in automatic
genre classification including structural features via parts-of-speech
 as well as lexical surface cues
\cite{petrenz2011stable}.

\textbf{Part of Speech Density.} Frequency of part-of-speech (POS)
tags, determined via the Stanford tagger, allows us to capture some of
the structural properties of text. For example, news statements can be
characterized by a high number of past tense verbs as well as proper
names.  We normalize the POS tag frequency w.r.t the sum of all
tags in a statement, to account for varying statement length.

\textbf{Verbs of Attribution and Quotation Marks.} News articles often
report statements by persons of repute, witnesses or other sources. We
approximate this by two features: Firstly, we count \emph{verbs of
  attribution} in $s$, via a list of 92 such verbs (\textit{claim,
  tell etc}) with POS tag \texttt{VB*} and normalize w.r.t the total
number of \texttt{VB*}. Secondly, we use \textbf{quotation marks} as a
potential indicator of \emph{paraphrasing}. The feature simply counts
the number of quotation marks in $s$, normalized w.r.t the statement
length.

\textbf{Temporal Proximity $\lambda(s)$.} Most Wikipedia statements
with news citations refer to relatively recent events, i.e. events
close to the time of the Wikipedia snapshot. Therefore, we use
\emph{temporal expressions} such as dates and years as distinguishing
features for news statements. We use a set of hand-crafted regular
expression rules to extract temporal expressions.\footnote{This proved
  to be more scalable than state-of-the-art extractors like
  HeidelTime~\cite{Strotgen:2010:HHQ:1859664.1859735} and Stanford's
  CoreNLP~\cite{DBLP:conf/acl/FinkelGM05} module}. We use the
  following rules: (1) \texttt{DD Month YYYY}, (2) \texttt{DD MM
    YYYY}, (3) \texttt{MM DD YY(YY)}, (4) \texttt{YYYY}, with
  different delimiters (whitespace, `-', `.'). We then compute
  $\lambda(s)=|Year(\mathbf{W}) - Year(s)|$.

\textbf{Discourse Analysis.} We use discourse connectives to annotate
the statements $s$ with \emph{explicit discourse relations} based on
an approach proposed by Pitler and
Church~\cite{DBLP:conf/emnlp/PitlerC09}. The annotations belong to the
categories $\{$\emph{temporal, contingency, comparison,
  expansion}$\}$, following the Penn Discourse Treebank annotation
\cite{prasad2008penn}. Some of the explicit discourse relations are
particularly interesting (i.e., \emph{temporal}) as they represent a
common language construct used in news articles that report event
sequences. The features are boolean indicators on whether $s$ contains
a specific explicit discourse relation.

\textbf{Language Model and Topic Model Scoring.} As surface lexical
features have been shown to be efficient in genre recognition
\cite{sharoff2010web}, we compute n--gram (up to n=3) language models
with \emph{Kneser-Ney} smoothing (LM) from news articles $N_t$ and
compute the score $\theta(s,N_t)$. This score shows how likely $s$ can
be constructed from the LM. Similarly, we compute topic models using
the LDA framework~\cite{DBLP:journals/jmlr/BleiNJ03}, where the score
is the jaccard similarity between $s$ and the topic terms from $N_t$.

\subsection{Entity-Structure Based Features}

Determining if a statement requires a news citation solely on language
style is not always feasible. We exploit the entity structure of $e$
and compute the probability of statements having a news citation given
its types $T(e)$ and sections $\Psi(e)$.

\textbf{Section-Type Probability.} A good indicator of the likelihood that a statement $s$ requires a news citation is the entity type it belongs to and the section type that it appears in. For instance, for type \texttt{Politician}, news statements have higher density in section \emph{`Early Life and Career'} as these tend to be more reflected in news. To avoid over-fitting we filter out entity types with fewer than 10 statements. Similarly, we filter out section with fewer than 10 statements, and in which they belong to the same citation category.

We compute the \emph{conditional probability} of having a news citation for $s$ for an entity type $t \in T(e)$ given a section type $\psi$.
\begin{equation*}\small
p(s=news|t,\psi) = \frac{\sum_{e\in \mathbf{W} \wedge t\in T(e)}\sum_{s \in S(e,\psi)}\1_{s \text{\texttt{ typeOf } news}}}{\sum_{e\in \mathbf{W}\wedge t\in T(e)}{|S(e,\psi)|}}
\end{equation*}

The $p(s=news|t,\psi)$ probability is likely to be a sparse feature, so we compute type and section news-priors. We compute section $p(s=news|\psi)$ and type news-priors $p(s=news| t)$ based on the news statement ratio that belong to a section or type, respectively.

Since $s$ is associated with an entity $e$, which has a set of types $T(e)$, we aggregate the computed type news-priors and the section-type joint probability into their \emph{min, max} and \emph{avg} probabilities.

\textbf{Type Co-Occurrence.} From the entity types $T(s)$ and $T(e)$ we measure the likelihood of type co-occurrence in news. The probability simply counts the co-occurrence between $t$ and $t'$ in
news statements with respect to their total co-occurrence. Examples of highly co-occurring types in news are \texttt{Politician} and \texttt{Organization}.

\begin{equation*}\small
p(s=news|t',t) = \frac{\sum_{e\in \mathbf{W}\wedge t\in T(e)}\sum_{s \in S(e)\wedge t'\in T(s)}\1_{s \text{\texttt{ typeOf } news}}}{\sum_{e\in \mathbf{W}\wedge t\in T(e)}\sum_{s \in S(e)}\1_{t'\in T(s)}}
\end{equation*}

\subsection{Learning Framework}\label{subsec:learning_framework}

\textbf{Learning Setup.} Wikipedia consists of a highly diverse set of
entities. A model trained on all entities is unlikely to
work.  For example, the types \texttt{Location} and
\texttt{Politician}
represent two highly divergent groups with regard to entity page
structure, the statements they contain and the way they are reported in news.

We therefore learn $SC$ for individual types in the YAGO type
taxonomy. The advantages of type specific functions $SC$ is that they
are trained on homogeneous entities, which helps the models predict
with greater accuracy. We take only types that have more than 1000
entity instances, resulting in 672 types. The types are organized from
very broad types such as  (\texttt{owl:Thing}) to very
specific types like \texttt{Serie\_A\_Players}.

To utilize the specialization and generalization in a principled
manner we transform the YAGO type taxonomy (DAG) into a hierarchical
DAG. This is utilized later on in order to find the right level of
type granularity for learning $SC$.

We assume that the hierarchy is rooted at \texttt{owl:Thing} and all
internal nodes are \emph{depth-consistent}, i.e. all paths from the
root to the node are of the same length. We obtain this by a simple
heuristic whereby for every \emph{child type} $\rightarrow$
\emph{parent type} we remove edges where the parent's \emph{depth
  level} in the taxonomy is higher than the \emph{minimum level} from
other parent nodes.

With this hierarchical type-taxonomy, we can determine the \emph{optimal level} of type granularity such that we have optimal performance in categorizing statements. For learning the type specific $SC$, we keep 10\% of entity instances for evaluation and the remainder for training. It is important to note that when we learn $SC$ for a given type, the training instances are sampled through \emph{stratified sampling} from all its \emph{children types}.

\textbf{Learning Model.} The functions $SC$ represent
\emph{multi-class} classifiers with classes corresponding to the
citation categories. Since we want to predict the news category
$c=$`\emph{news}' with high accuracy, one question is why we do not
pose this as a \emph{binary} classification problem, where a statement
is categorized as news or not. We used the \emph{multi-class}
classifiers because they give us a more balanced distribution
when compared to merging all non-news statements into
a single category.

Finally, we opt for \emph{Random Forests} (RF)~\cite{Breiman2001} as
our supervised machine learning model. We experimented with other
models, but the differences in performance are marginal, and RF have
superior learning time. We train the models on the full feature set in
Table~\ref{tbl:sc_feature_list}.

\section{Citation Discovery}\label{sec:missing_citations}

For the citation discovery task, we follow the \emph{citation
  policy}\footnote{\small{\url{https://en.wikipedia.org/wiki/Wikipedia:Citing_sources}}}
guidelines in Wikipedia and single out three key properties on what
makes a \emph{good citation}.

\begin{enumerate}
\itemsep0em
  \item the statement should be \emph{entailed} by the cited  news article
  \item the statement should be \emph{central} in the cited news article
  \item the cited news article should be from an \emph{authoritative} source 
\end{enumerate}

We approach the citation discovery for news statements as
follows. We use statement $s$ as a query (see
Section~\ref{subsec:query_construction}) to retrieve the top--$k$ news
articles from $\N$ as citation candidates for $s$. We then classify
the candidate citations as either `\emph{correct}' or
`\emph{incorrect}', depending on whether they meet the above criteria of a
\emph{good citation}.

In order to do so, we compute features for each pair $\langle s,
n_i\rangle$, w.r.t the individual sentences of a news article
$n_i$. The feature vectors become the following $\langle s,
[\sigma_i^1, \sigma_i^2,\ldots, \sigma_i^j]\rangle$, where
$\sigma_i^j$ represents the $j$-th sentence from $n_i$.

Since the number of sentences $\sigma_i$ varies across news articles,
we aggregate the individually computed features at sentence level into
the corresponding \emph{min, max}, \emph{average}, \emph{weighted
  average}, and \emph{exponential decay function} scores as shown
below.
\begin{equation*}\small
\langle s, \min_{j}F(\sigma_i^j), \max_{j}F(\sigma_i^j), Avg(F(\sigma_i)), \sum_{\sigma_i^j}\frac{1}{j}* F, \sum_{\sigma_i^j} F^{\frac{1}{j}},\ldots\rangle
\end{equation*}
where $F$ is a feature from the complete feature list in Table~\ref{tbl:missing_citations_feature_list}. 

\begin{table*}[ht!]\small
\centering
\scalebox{0.9}{
\begin{tabular}{p{1.1cm} p{5cm} p{1.1cm} p{4.4cm} p{1cm} p{3cm}}
\toprule
\multicolumn{2}{c}{\emph{entailment}} & \multicolumn{2}{c}{\emph{centrality}} & \multicolumn{2}{c}{\emph{news-domain authority}} \\
\midrule
$\mathbf{J}^{n}(s, \sigma_i^j)$ & n--gram overlap between $s$ and $\sigma_i^j$ and similarity headline of $n_i$ and $s$  & 
$\mathbf{J}(s, \sigma_i^c)$ & jaccard similarity between $s$ and central sentence $\sigma_i^c$  & 
$p(\mathbf{D}[n_i]|t)$ & domain authority of news article $n_i$ for type $t$ \\

$\mathbf{J}^{P}(s, \sigma_i^j)$ &  NNP phrase overlap between $s$ and $\sigma_i^j$ &  
$\mathbf{J}^{P}(s, \sigma_i^c)$ &  NNP phrase overlap between $s$ and $\sigma_i^c$ & 
$p(\mathbf{D}[n_i]|s)$ & domain authority of $n_i$ for section $s$  \\

$\theta^1(s,\cdot)$ & $s$ unigram LM from news article $n_i$ and n--gram LM from news articles in $N_t$ & 
$\mathbf{J}^{n}(s, \sigma_i^c)$ & n--gram overlap between $s$ and $\sigma_i^c$  &\\

$K(s, \sigma_i^j)$ & tree kernel similarity between $s$ and $\sigma_i^j$   & 
$K(s, \sigma_i^c)$ & tree kernel similarity between $s$ and $\sigma_i^c$ &\\

$LDA(s,N_t)$ & term overlap between $s$ and topic terms from news in $N_t$  & 
$\phi(e,n_i)$ & the relative entity frequency of $e$ in $n_i$ &  \\

$freq(e)$ & occurrence frequency of $e$ in the title and body of $n_i$ & 
$\phi(\gamma(s),n_i)$ & relative entity frequency of $e\in \gamma(s)$ in $n_i$ & & \\

\emph{baseline features} & retrieval score from the IR model for $n_i$ and its rank  & & & & \\
\bottomrule
\end{tabular}}
\caption{\small{Extracted feature set for the citation discovery task.}}
\label{tbl:missing_citations_feature_list}
\end{table*}

\subsection{Query Construction}\label{subsec:query_construction}

We use the \emph{statement text} as query which can vary from a sentence to a paragraph.

One way to improve the likelihood of obtaining good citation
candidates from top--$k$ articles is through \emph{query construction}
approaches (QC). It has been shown that in similar cases
where the query corresponds to a sentence or paragraph, QC approaches
are necessary to increase the accuracy of IR models. Henzinger et
al.~\cite{DBLP:conf/www/HenzingerCMB03} propose several QC approaches
that weigh query terms based on the \emph{tf--idf} score.

We experimented with different QC approaches from
\cite{DBLP:conf/www/HenzingerCMB03} and their impact on finding news
articles in $\mathcal{N}^W$. We found that \emph{QCA1Base} performed
best and use it in the remainder of the paper. In \emph{QCA1Base}, the terms
extracted from the statements are weighted based on \emph{tf--idf},
with $tf$ and $idf$ are computed w.r.t the other statements under
consideration.

In principle, one should consider all retrieved articles from the
result set. However, this is not only computationally expensive for
our subsequent learning step but also unbalances our training set. To
determine a reasonable retrieval depth, we experimented with 1000
randomly chosen statement queries with QC and determined the hit-rate
at retrieval depth $k$ , i.e.  whether the cited article is
retrieved in the top---$k$ articles.

\begin{figure}[ht!]
\centering
\includegraphics[width=0.7\columnwidth]{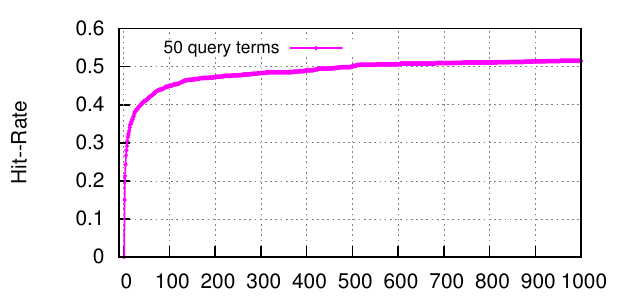}
\caption{\small{Hit-rate of articles in $\N^W$ up to rank 1000 (x--axis) for 1000 news statements, respectively \emph{QCA1Base} queries.}}
\label{fig:index_coverage}
\end{figure}

Figure~\ref{fig:index_coverage} shows the hit-rate in top--1000 with
top 50 ranked query terms and with \emph{divergence from randomness}
query similarity measure~\cite{Amati:2002:PMI:582415.582416} for our
\emph{random sample} of 1000 news statements.

We focus on the top--100 retrieved news articles as potential
citations for $s$, as the achieved hit-rate beyond the top--100 shows
only minor improvement. In Figure~\ref{fig:index_coverage}, we also
note that the hit-rate does not go beyond 50\%. We found that most of
the news articles that are not retrieved are either missing or
non-English articles in $\N^W$.

\subsection{Textual Entailment Features}\label{subsec:entailment}

As the citation is supposed to give close evidence for the statement's
content, in the ideal case the cited news article should fully
\textit{entail} the statement, i.e. the statement should be derivable
from the news article. The recognition of textual entailment has been
the study of extensive research in the last 10 years; cf
\cite{dagan2013recognizing} for an overview. A full treatment of
entailment needs extensive world knowledge and inference rules; we
here restrict ourselves to much simpler lexical and syntactic
similarity methods used in baseline entailment systems and leave the
extensions to future work.\footnote{\small{Off-the-shelf entailment
    systems exist but are too slow to use at scale.}}

\textbf{IR Baseline Features.} We use the retrieval model as a pre-filter
to find candidate news articles as citations for $s$. The retrieval
model also provides us with two possible features for the learning
model: firstly, a \emph{matching score} of $n_i$ for query $s$, where
the score corresponds to the \emph{divergence from randomness} query
similarity measure~\cite{Amati:2002:PMI:582415.582416}.
Secondly, the retrieval rank of $n_i$. We use the IR model as our
\emph{baseline} and hence refer to them as \emph{baselines features}.

\textbf{Tree Kernel Similarity.} Lexical similarity measures in many
cases fail to capture the joint \emph{semantic} and \emph{syntactic}
similarity. For this purpose, we consider the \emph{tree kernel}
similarity measure proposed in~\cite{DBLP:conf/emnlp/Kate08}. We
first compute the \emph{dependency parse trees} of $s$ and
$\sigma_i^j$ using the Stanford tagger~\cite{toutanova2000enriching}, and then compute the tree
kernel, $K(s, \sigma_i^j)$. Tree kernel similarity through the
dependency parse tree measures the maximum matching subtrees between
$s$ and $\sigma_i^j$, where the matching subtrees have the same
syntactic and semantic meaning. We refer the reader to
\cite{DBLP:conf/emnlp/Kate08} for details.

\textbf{LM \& Topic Model Scoring.} From an article $n_i$ we compute a
\emph{unigram LM} and compute $\theta(s, n_i)$ as the likelihood of
$s$ being generated from the computed LM.
In addition, we compute \emph{n--gram} LM (with $n$ up to 3) from articles in $N_t$, and compute the score $\theta^n(s, N_t)$ accordingly. 

Similarly, we compute \emph{LDA topic
  models}~\cite{DBLP:journals/jmlr/BleiNJ03} for entity types,
specifically from articles in $N_t$. This follows the intuition that
content usually is clustered around specific topics, i.e. for type
\texttt{Politician} most discussions are centered around politics,
career, etc. The topic score is the Jaccard similarity between $n_i$ and
the topic terms.

\subsection{Centrality Features}\label{subsec:centrality}

\textbf{Similarity to most central news sentence.} As described above
  we compute similarity features between $s$ and sentences in $n_i$.
  However, some sentences in $n_i$ are more central than others.
  Hence, the computed features between the pairs $\langle s,
  [\sigma_i^1, \sigma_i^2,\ldots, \sigma_i^j]\rangle$, do not have
  uniform weight. Therefore, we find the most \emph{central sentence}
  $\sigma_i^c$ in $n_i$ and distinguish the computed
  entailment/similarity features between $s$ and $\sigma_i^c$.

We compute centrality of a sentence in $n_i$ through the TextRank
approach introduced in~\cite{DBLP:conf/emnlp/MihalceaT04}. We first
construct a graph $G=(V,E)$ from $n_i$, where $V$ corresponds to the
sentences of $n_i$, with edges in $E$ weighted with the Jaccard
similarity between any two sentences, in this case $\sigma_i^j \in V$.
Computation of centrality for any vertex $\sigma_i^j$ is similar to
that of PageRank, with slight changes accounting for the weighted
edges between vertices.
\begin{equation}\small
\Gamma(\sigma_i) = (1-d) + d * \sum\limits_{\sigma_j \in In(\sigma_i)}\frac{\mathbf{J}(\sigma_i, \sigma_j)}{\sum\limits_{\sigma_k\in Out(\sigma_j)}{\mathbf{J}(\sigma_j, \sigma_k)}} \Gamma(\sigma_j)
\end{equation}
where $d$ is the damping factor ($d=0.85$), a common value in PageRank computation. The computation converges if the difference in the score of $\Gamma(\sigma_j)$ in two consecutive iterations is small.

\textbf{Relative Entity Frequency.} The importance of $e$ in $n_i$ is
crucial when finding citations for $s$. This importance is partially
mirrored simply in how often $e$ is mentioned in $n_i$. However,
another genre-typical property of news is its inverted pyramid
structure, i.e. the most important information is mentioned at the
beginning of the article. We therefore 
measure relative entity
frequency of $e$ in $n_i$ based on an approach described
in~\cite{DBLP:conf/cikm/FetahuMA15}. It attributes higher weight to
entities appearing in the top paragraphs of $n_i$, where the weight
follows an exponential decay function.

\begin{equation}\label{eq:rel_freq}\small
\phi(e,n) =  \frac{|\rho(e,n)|}{|\rho(n)|}\sum\limits_{\rho\in \rho(n)}\left(\frac{tf(e,\rho)}{\sum\limits_{e'\neq e}tf(e',\rho)}\right)^{\frac{1}{\rho}}
\end{equation}
where $\rho$ represents a news paragraph from $n$ and $\rho(n)$
indicates the set of all paragraphs. $tf(e,\rho)$ indicates the
frequency of $e$ in $\rho$. With $|\rho(e,n)|$ and $|\rho(n)|$ we
indicate the number of paragraphs in which entity $e$ occurs and the
total number of paragraphs.

Additionally we consider the relative entity frequency for entities in
$e\in\gamma(s)$ and measure the \emph{minimum, maximum} and
\emph{average} relative entity frequency scores.

\subsection{News-Domain Authority Features}\label{subsec:authority}

Wikipedia's editing policy distinguishes clearly between more and
less-established news outlets and prefers the former (see the
Introduction).  We therefore compute the authority of news domains
w.r.t entity types and sections.

We will
denote the \emph{domain} of the news article referred from $s$ as
$D[s]$, and with $D$ any arbitrary domain.

\textbf{Type-Domain Authority.} Authority of news domains is
non-uniformly distributed across types. For types such as  \texttt{Politician}
the authority of domains like \emph{BBC} is higher than for types such as 
\texttt{Athletes}, where a domain specialized in sports news is more
likely to be  authoritative. We capture the \emph{type-domain
  authority} as follows:
\begin{equation*}\small
p(D|t) = \frac{\sum_{e\in\mathbf{W}\wedge t\in T(e)}\sum_{s\in S(e)}\1_{D=D[s]}}{\sum_{e\in\mathbf{W}\wedge t\in T(e)}\sum_{s \in S(e)}D[s]}
\end{equation*}

\textbf{Section-Domain Authority.} We measure the authority of domains
associated to certain entity sections. The density of news references
across sections varies heavily. Therefore, it is natural to consider the
authority of news domains for a given section.
\begin{equation*}\small
p(D|\psi) = \frac{\sum_{e\in\mathbf{W}}\sum_{s \in S(e,\psi)}\1_{D=D[s]}}{\sum_{e\in\mathbf{W}}\sum_{s \in S(e,\psi)}D[s]}
\end{equation*}

Note that these features compute news outlet authority with regard to
current Wikipedia usage, which we seek to re-create. An alternative we
intend to look at in future work is to measure authoritativeness via
Wikipedia-external measures of news outlets, such as page visits or
interlinkage.

\section{Statement Categorization\\ Evaluation}\label{sec:t1_results}

Here we describe the evaluation of our approach for SC. Since we
consider a type taxonomy, we have a hierarchy of models. Each
statement belongs to an entity, which in turn is a child to a
type (node) in the hierarchy.  Consequently, we construct each model
from training instances (statements) that are its children. We focus
on two aspects of our approach (i) performance of models at varying
depths, and (ii) performance of various feature classes.

We provide the detailed results for the statement categorization task
and the corresponding ground-truth data at the paper
URL\footnote{\label{paper_url}\url{http://l3s.de/~fetahu/cikm2016/}}.

\subsection{Experimental Setup}\label{subsec:t1_exp_setup}

\textbf{Setup.} We consider 672 entity types from our Yago taxonomy,
for which we learn individual $SC$ models. We consider types that have
more than 1000 entity instances. The level of granularity in the YAGO
taxonomy has a maximum depth of 20, while the root type is
\texttt{owl:Thing} containing all possible entities.

\textbf{Train/Test.} We learn the $SC$ models using up to 90\% of the
entity instances of a type $t$ as training set, and the remainder of
10\% for evaluation. We use \emph{stratified sampling} to pick
entities of type $t$ and its subtypes for the train and test set. We
train and test $SC$ models over 6 million statements coming from 1.3
million entities.

\textbf{Metrics.} We evaluate the performance of $SC$ with
\emph{precision} $P$, \emph{recall} $R$ and $F1$. A statement is
categorized correctly if the predicted category corresponds to the
ground-truth.

\subsection{Results and Discussion}\label{subsec:t1_results}

The following discussion focuses on the results for the statement
categorization task for the \texttt{news} category. Due to space
constraints we report the first three type levels in the Yago
taxonomy, specifically the immediate child \texttt{Legal Actor Geo} of
\texttt{owl:Thing}.\footnote{For readability we remove the
    \emph{wordnet} prefix from the types and their numerical ID
    values.} The results for the remainder of the types are
accessible at the URL\footnoteref{paper_url}.

Table~\ref{tbl:yagoLegalActorGeo_t1} shows the results for $SC$ models evaluated over 61k entities and trained with up to 550k entities, depending on the training sample size $\tau \in [1\%,90\%]$.  The results for this type represent more than 47\% of the total set of entities in our evaluation dataset.

The overall performance of $SC$ for all types for $\tau=90\%$ measured through \emph{micro-average} precision is 0.57. Since a statement belongs to multiple types $T(s)$, we decide the category of $s$  based on majority as categorized from the individual $SC$ models.

\begin{table*}[ht!]\small
\centering
\scalebox{0.9}{
\begin{tabular}{p{0.8cm} l p{3cm} P{0.6cm} P{0.6cm} P{0.6cm}  P{0.5cm} P{0.6cm} P{0.6cm}  P{0.6cm} P{0.6cm} P{0.6cm} }
\toprule
\multicolumn{12}{c}{\texttt{yagoLegalActorGeo}} \\
\midrule
\texttt{Level} & \texttt{Parent Type} & \texttt{Child Type} & \multicolumn{3}{c}{$1 \leq \tau \leq 10$} & \multicolumn{3}{c}{$10 < \tau \leq 50$} & \multicolumn{3}{c}{$50 < \tau \leq 90$}\\
\midrule
 & & & \textbf{P} & \textbf{R} & \textbf{F1} & \textbf{P} & \textbf{R} & \textbf{F1} & \textbf{P} & \textbf{R} & \textbf{F1}\\
\cmidrule{4-12}

\texttt{L.0} & \texttt{owl:Thing} & \texttt{Legal Actor Geo} & 0.48 & 0.36 & 0.41 & 0.51 & 0.43 & 0.47 & 0.53 & 0.47 & 0.50\\

\hline

\multirow{2}*{\texttt{L.1}} & \multirow{2}{2.4cm}{\texttt{Legal Actor Geo}} & \texttt{Legal Actor} & 0.51 & 0.34 & 0.41 & 0.54 & 0.41 & 0.47 & \textbf{0.56} & \textbf{0.45} & \textbf{0.50}\\
& & \texttt{location} & 0.30 & 0.29 & 0.29 & 0.34 & 0.40 & 0.37 & 0.36 & 0.45 & 0.40\\

\hline

\multirow{3}*{\texttt{L.2}} & \multirow{2}*{\texttt{location}} & \texttt{region} & 0.30 & 0.28 & 0.29 & 0.35 & 0.40 & 0.37 & 0.37 & 0.44 & 0.40\\
& & \texttt{point} & 0.30 & 0.1 & 0.14 & 0.38 & 0.22 & 0.28 & 0.39 & 0.26 & 0.32\\[1.5ex]

& \texttt{Legal Actor} & \texttt{person} & 0.53 & 0.36 & 0.43 & 0.56 & 0.43 & 0.49 & \textbf{0.58} & \textbf{0.46} & \textbf{0.51}\\

\hline

 \multirow{5}*{\texttt{L.3}} & \multirow{5}*{\texttt{person}} & \texttt{preserver} & 0.63 & 0.31 & 0.42 & 0.67 & 0.46 & 0.54 & \textbf{0.67} & \textbf{0.49} & \textbf{0.57}\\
& &  \texttt{authority} & 0.53 & 0.20 & 0.29 & 0.62 & 0.24 & 0.35 & 0.65 & 0.33 & 0.44\\
& &  \texttt{contestant} & 0.59 & 0.43 & 0.50 & 0.62 & 0.52 & 0.57 & 0.64 & 0.56 & 0.60\\
& &  \texttt{leader} & 0.53 & 0.26 & 0.34 & 0.59 & 0.34 & 0.43 & 0.61 & 0.37 & 0.46\\
& &  \texttt{wc Living people} & 0.55 & 0.37 & 0.44 & 0.58 & 0.44 & 0.50 & 0.59 & 0.47 & 0.52\\

\bottomrule
\end{tabular}}
\caption{\small{Results for the \emph{statement classification} for
    entities of type \texttt{yagoLegalActorGeo}. Results are
    aggregated for the different sample ranges $\tau$ and shown
    at different levels of entity types in the YAGO type hierarchy.}} 
\label{tbl:yagoLegalActorGeo_t1}
\end{table*}

\subsubsection{Level of Type Granularity}

As expected, we observe that model performance depends on the type
level (cf. Table~\ref{tbl:yagoLegalActorGeo_t1}). A unified model from
heterogeneous training instances performs
poorly:  the $SC$ model for the main type \texttt{Legal
  Actor Geo} achieves a precision P=0.527 with high variance across
its subtypes. Comparing the types at depth level 3, the difference in
terms of precision can go as high as 15\% between \texttt{Legal Actor
  Geo} and the best performing subtype \texttt{preserver}.

At higher depths, performance of the $SC$ models often improves
significantly as the instances belonging to a given type become more
homogeneous. For example, the fine grained entity type \texttt{wcat
  Italian footballers} has a precision of P=0.87 and recall of R=0.58,
which constitutes a 50\% precision and a 26\% recall improvement over
its parent type \texttt{Person}. However, the performance improvement is
not monotonically increasing. In some fine-grained types, there is in
fact a performance reduction which can be attributed to
over-fitting. This suggests that there is indeed a sweet spot in terms
of choice of the best performing model for an instance. We observed
that the instances that are children of \texttt{person} 
showed best performances between levels 5 and 8.

Our models perform poorly for types such as \texttt{location} since 
 location pages have a lower news density. We again observe that news
articles are usually centered around people and its instances benefit
the most from our approach. We also observe that the performance of
our approach is sensitive to the type hierarchy. The choice of YAGO
as a taxonomy is due its fine-grained types. However, there exist many
long-tail entities that are direct descendants from the higher levels
and fail to leverage the homogeneity of fine-grained types. We also 
perform poorly on such instances. 

In the YAGO taxonomy, the entities are distributed normally with a
mean at depth level 8, which contains around 36\% of entities. The
long tail with types lower than depth level 8 accounts for 28\% of
entities in the YAGO taxonomy.

We focussed on the category \texttt{news} in our discussion and in
Table~\ref{tbl:yagoLegalActorGeo_t1}.  Performance of $SC$ models for
the categories $c=\{$\texttt{web}, \texttt{book}, \texttt{journal}$\}$ and type
\texttt{person} is P=0.62 and R=0.59, P=0.29 and R=0.69, and P=0.25
and R=0.26, respectively. The relatively high score for the
\texttt{web} category can be attributed to the high density of
statements of category \texttt{web}, accounting for more than 54\% of
the total statements. Hence, by always choosing \texttt{web} as the
category of a statement we get an average precision of 0.54.

\subsubsection{Convergence and Feature Ablation}

\begin{figure}[ht!]
\centering
\includegraphics[width=0.8\columnwidth]{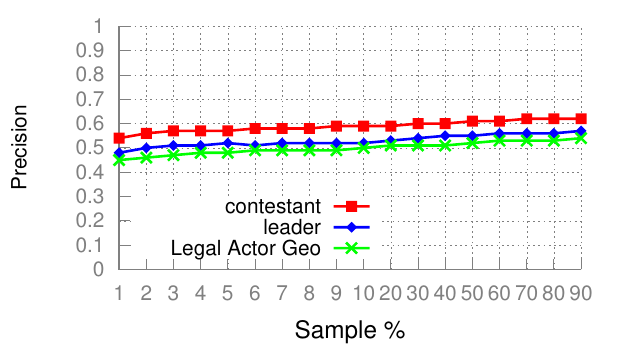}
\caption{\small{Learning curve for $SC$ measured for different sample sizes.}}
\label{fig:t1_learning_curve}
\end{figure}

\textbf{Convergence.} We measure the amount of training data required for the models to converge to optimal performance. Figure~\ref{fig:t1_learning_curve} shows the learning curve for some of the types reported in
Table~\ref{tbl:yagoLegalActorGeo_t1}. We see that $SC$ models converge and achieve optimal performance early on with a sample around 7\% to 10\%.

\begin{figure}[h!]
\centering
\includegraphics[width=0.8\columnwidth]{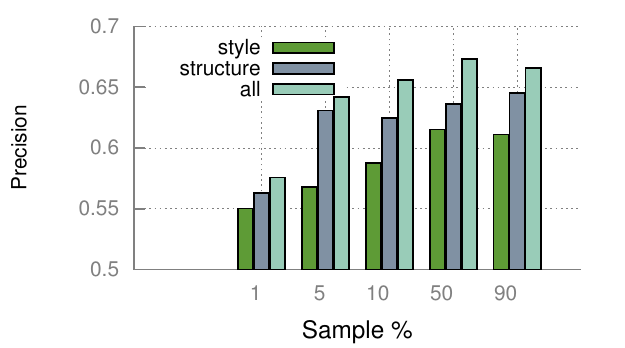}
\caption{\small{Feature ablation for features in $SC$ for type \texttt{preserver}}}
\label{fig:feature_ablation_yagoLegalActorGeo}
\end{figure}

\textbf{Ablation.}  We apply a feature ablation test for the different
different features groups from Table~\ref{tbl:sc_feature_list}.
Figure~\ref{fig:feature_ablation_yagoLegalActorGeo}
shows the results for the feature groups \emph{language style}, and
\emph{entity structure}. The highest gain is achieved with the feature
group \emph{entity structure}, which reveals the challenging nature of
the task where language style features cannot be applied alone.

\section{Citation Discovery Evaluation}\label{sec:t2_evaluation}

In this section, we evaluate the \emph{citation discovery} task for
\emph{news statements}. We perform an extensive evaluation for
approximately 22k news statements and discover citations from a
real-word news collection with 20 million articles in a timespan of
two years.

\subsection{Statement and News Collection}\label{subsec:data_gdelt_news}

We limit ourselves to the subset of news statements with citations
to news articles in $\N^W$ from 2013 to 2015. The resulting set contains 22k
news statements with 27k news article citations in
$\N^W$.\footnote{\small{A statement can have more than one citation.}}
We denote this temporal slice of news articles in $\N^W$ by $\N^W_{13-15}$.

As finding the right citation from this preselected collection is
easier than the realistic scenario of finding a citation among all
possible news, we also collected all English news articles from the
period [2013-08, 2015-08] from the GDelt
project\footnote{\small{\url{http://gdeltproject.org/}}}.  We call the
resulting high-coverage dataset $\N^G$.

We merge $\N^G$ with $\N^W_{13-15}$ and call the resulting dataset
$\N = \N^W_{13-15} \cup \N^G$. The set $\N$
contains around 20 million news articles.  $\N^W_{13-15}$ accounts for
less than 1\% in $\N$, making the correct articles hard to find.

\subsection{Evaluation Strategies}\label{subsec:eval_strategy}

\textbf{Evaluation Strategy \texttt{E1}:} In this scenario, we, for
each news statement $s$, only consider the pairs $\langle s,
n\rangle$, where $n\in N_s$ as correct and all other possible
citations as incorrect. This allows for fully automatic evaluation but
is only a \emph{lower bound} for $FC$, as there can be additional
articles that are relevant for $s$ but do not exist in $N_s$.  We
therefore also consider a variant \texttt{E1+FP}, where we consider
$n'\notin N_s$ as additional correct citations if the
similarity (based on the \emph{jaccard} similarity) to one of the articles in $N_s$ is above 0.8.

\textbf{Evaluation Strategy  \texttt{E2}:}
\texttt{E2} assesses the
true performance of $FC$. In this case, apart from already existing
citations for $s$ from $N_s$, we assess through \emph{crowd-sourcing}
the appropriateness as citations of articles $n\in \N \wedge n \notin
N_s$.

We set up the crowd-sourcing experiment for \texttt{E2} as
follows. For a statement $s$ and an article $n_i\notin N_s$ marked as
correct by $FC$, we ask the crowd to compare $n_i$ with the
ground-truth article $n \in N_s$ and answer the question \emph{`Which
  of the two shown news articles is an appropriate citation for the
  statement?'}. The workers are shown $s$ as well as $n_i$ and the
ground truth article in random order without an indication which one
is the ground truth.  We provide the following response options: (i)
\emph{first}, (ii) \emph{second}, (iii) \emph{both}, (iv) \emph{none},
and (v) \emph{insufficient info}. We deployed the experiment in
CrowdFlower\footnote{\small{\url{https://www.crowdflower.com}}} and
chose only high quality workers to ensure the reliability of our
experiments\footnote{We select workers with the highest quality as
  provided by the CrowdFlower platform.}. Furthermore, we removed workers who did not spend the minimum amount of two minutes to assess the appropriateness of a citation\footnote{The amount of two minutes was decided based on the number of citations the workers had to assess per page (consisting of 5 citations to assess).}.

We collect three judgments per question. We count citations as correct
which are ground-truth articles or articles which the majority
of workers judge as appropriate citations.

\subsection{Experimental Setup}\label{subsec:t2_exp_setup}

\textbf{Retrieval model.} We use the retrieval model in
  \cite{Amati:2002:PMI:582415.582416} via the implementation provided
  by Solr\footnote{\small{\url{http://lucene.apache.org/solr/}}}. We
  use the top-100 retrieved news articles for a statement as candidate citations, from which we perform feature extraction and learn our SC models.

\textbf{Learning Setup.} We learn classifiers specific to entity types
for a total of 83 types. We limit ourselves to types that have news
statements in the date range 2013-2015 and with at least 100 entity
instances. From our set of 22k statements, we randomly sample
statements from each entity type if they have more than 1000
instances, otherwise we take all statements. Training and testing data
consist of the pairs $\langle s, n_i\rangle$, where $s$ is a news
statement, and $n_i$ is one of the top--100 citation candidates which
we retrieve from $\N$. We split training and testing data per
statement $s$, where each $s$ and all its candidates are included
\emph{completely} either in the training or test set.

\textbf{Learning Approach.}  We learn the $FC$ models as supervised
binary classification models using random forests
RF\cite{Breiman2001}.  We predict $\langle s,
n\rangle\in$\emph{`correct', `incorrect'}, i.e. if a candidate news
article is an appropriate citation for $s$ or not. We optimize for the
\emph{`correct}' class. The correct labels in training and automatic
evaluation $E1$ are all part of $\N^W_{13-15}$, which makes up less than
1\% of our news collection $N$. Therefore, we learn $FC$ as a
\emph{cost-sensitive} classifier.

\textbf{Metrics.} We evaluate performance of $FC$ models
via precision $P$, recall $R$, and $F1$ score.

\textbf{Baselines.} We consider two baselines (\textbf{B1} and
\textbf{B2}) for this task. For $\mathbf{B1}$, we use the
\emph{divergence from randomness}
model~\cite{Amati:2002:PMI:582415.582416} to retrieve news articles
from $\N$ for $s$ and simply  suggest the top--1 article as citation. In
$\mathbf{B2}$ we learn a supervised model based on the \emph{IR baseline
  features} (see Table~\ref{tbl:missing_citations_feature_list}).

\subsection{Results and Discussion}\label{subsec:t2_results}

Table~\ref{tbl:t2_results_e1} shows the results for all evaluation
strategies for the citation discovery task.  We only display detailed
results for the top--10 best performing entity types out of the 83
types in our evaluation.  The results in each row in
Table~\ref{tbl:t2_results_e1} show the best performance we achieve for
the individual types, while varying the variables such as the
\emph{training sample size} and \emph{feature number}. We show results
with a maximum of 60\% training sample size.

We report additionally the overall performance of $FC$ models across
all 83 types through \emph{micro-average} in the last row in
Table~\ref{tbl:t2_results_e1}. The detailed results are accessible at
the paper URL\footnoteref{paper_url}.

\subsubsection{E1: Automated Evaluation}\label{subsec:e1_t2_results}

In Table~\ref{tbl:t2_results_e1}, in the third column, we show the evaluation results for the strategy \texttt{E1}. 

Results for \texttt{E1} are encouraging given the fact that in
top--100 news candidates retrieved from $\mathcal{N}$ only 1\% of the
news are `\emph{correct}' (on average one relevant citation in
$\N^W_{13-15}$ per statement). Furthermore, as shown in
Figure~\ref{fig:index_coverage} the highest recall we get at top--100
is on average around 45\%.

We achieve the best performance in terms of precision for the entity
type \texttt{football} \texttt{player}, with precision P=0.80 and a
recall of R=0.30. For F1 the best performing type in this setup is the
entity type \texttt{player} with F1=0.57.

Using the evaluation strategy \texttt{E1+FP}, we consider as relevant
all \emph{false positive} (FP) articles which are highly similar to
the ground-truth articles $N_s$ (above 0.8 similarity). Even though
the FP articles do not exist in our ground-truth, the high similarity
to the ground-truth article is a  strong indicator for them being
relevant citations. Using this strategy, the results improve for some
of the types with up to 8\% in terms of precision. For type
\texttt{entertainer} we have an increase of 11\%. In absolute numbers,
by considering the highly-similar FP articles as relevant we gain an
additional 757 news articles out of 12,877, i.e.  an additional 6\%
news citations.

\begin{table*}[h!]\small
\centering
\scalebox{0.9}{
\begin{tabular}{l P{0.5cm} P{0.5cm} P{0.5cm} P{0.5cm} P{0.5cm} P{0.5cm} | P{0.5cm} P{0.5cm} P{0.5cm} | l  | l | P{0.5cm} P{0.5cm}}

\toprule
& \multicolumn{3}{c}{\texttt{B1}} &  \multicolumn{3}{c}{\texttt{B2}} & \multicolumn{3}{c}{\texttt{E1}} & \multicolumn{1}{c}{\texttt{E1 + FP}}  & \multicolumn{1}{c}{\texttt{E2}}  & & \\
\cmidrule{2-14}
\texttt{type} & \textbf{P} & \textbf{R} & \textbf{F1} & \textbf{P} & \textbf{R} & \textbf{F1}&  \textbf{P} & \textbf{R} & \textbf{F1} & \multicolumn{1}{c|}{\textbf{P}} & \multicolumn{1}{c|}{\textbf{P}}  & \#feat. & \%train\\
\cmidrule{1-14}
\texttt{player} & 0.37 & 0.36 & 0.37 & 0.31 & 0.28 & 0.29 & 0.67 & 0.46 & 0.55 & 0.71 $\blacktriangle$ (5.63\%) & \textbf{0.85} $\blacktriangle$ (21.18\%) & 20 & 60\\
\texttt{entertainer} & 0.32 & 0.31 & 0.31 & 0.16 & 0.18 & 0.17 & 0.70 & 0.33 & 0.45 & 0.78 $\blacktriangle$ (10.26\%) & \textbf{0.90} $\blacktriangle$ (22.22\%) & 40 & 60\\
\texttt{causal agent} & 0.26 & 0.26 & 0.26 & 0.17 & 0.21 & 0.19 & 0.73 & 0.28 & 0.41 & 0.77 $\blacktriangle$ (5.19\%) & \textbf{0.88} $\blacktriangle$ (17.05\%) & 40 & 60\\
\texttt{location} & 0.21 & 0.19 & 0.20 & 0.21 & 0.23 & 0.22 & 0.55 & 0.26 & 0.35 & 0.62 $\blacktriangle$ (11.29\%) & \textbf{0.83} $\blacktriangle$ (33.73\%) & 30 & 60\\
\texttt{artist} & 0.31 & 0.31 & 0.31 & 0.24 & 0.27 & 0.25 & 0.67 & 0.21 & 0.32 & 0.67 & \textbf{0.85} $\blacktriangle$ (21.18\%) & 50 & 60\\
\texttt{football player} & 0.31 & 0.31 & 0.31 & 0.29 & 0.38 & 0.33 & 0.80 & 0.30 & 0.43 & 0.80 & \textbf{0.90} $\blacktriangle$ (11.11\%) & 50 & 60\\
\texttt{wcat Living people} & 0.27 & 0.26 & 0.26 & 0.21 & 0.18 & 0.2 & 0.67 & 0.23 & 0.34 & 0.70 $\blacktriangle$ (4.29\%) & \textbf{0.85} $\blacktriangle$ (21.18\%) & 50 & 50\\
\texttt{creator} & 0.34 & 0.32 & 0.33 & 0.25 & 0.24 & 0.24 & 0.74 & 0.25 & 0.38 & 0.74 & \textbf{0.91} $\blacktriangle$ (18.68\%) & 50 & 50\\
\texttt{organism} & 0.29 & 0.28 & 0.28 & 0.22 & 0.19 & 0.2 & 0.69 & 0.30 & 0.41 & 0.70 $\blacktriangle$ (1.43\%) & \textbf{0.83} $\blacktriangle$ (16.87\%) & 40 & 60\\
\texttt{person} & 0.26 & 0.24 & 0.25 & 0.21 & 0.23 & 0.22 & 0.64 & 0.35 & 0.46 & 0.66 $\blacktriangle$ (3.03\%) & \textbf{0.85} $\blacktriangle$ (24.71\%) & 20 & 60\\
\midrule
\emph{micro-average} & 0.25 & & & 0.21 & & \multicolumn{1}{c}{} & 0.67 & & \multicolumn{1}{c}{} & \multicolumn{1}{l}{0.71 $\blacktriangle$ (5.6\%)} & \multicolumn{1}{l}{\textbf{0.86 $\blacktriangle$ (22.00\%)}}\\
\bottomrule
\end{tabular}}
\caption{\small{Top--10 best performing entity types for the $FC$ task. \texttt{E1+FP} and \texttt{E2} columns show the improvement for P over \texttt{E1}. Right most column shows the configuration with which we learn the $FC$ models. The last row shows the \emph{micro-average} precision across all $FC$ models.} }
\label{tbl:t2_results_e1}
\end{table*}

Baselines \textbf{B1} and \textbf{B2} show the difficulty of the
citation discovery task. In particular, we show that standard IR
models struggle with this task. Choosing only the top--1 article for
citation (\textbf{B1}) achieves only up to P=0.37. On the other hand, for \textbf{B2}, we see
that we cannot learn well using only the IR baseline features, and perform even worse than using \textbf{B1}.

\subsubsection{E2: Automated+Crowdsourced Evaluation}\label{subsec:e2_t2_results}

For \texttt{E2}, we report results after re-evaluating performance of
$FC$ models via gathering judgements for false positive (FP) news
articles suggested as citations for $s$. We evaluate 11,803 false
positive news article citation candidates for the top--10 entity types
in Table~\ref{tbl:t2_results_e1}, from 6.9k news statements. As
reported above, crowd-workers could choose between both ground truth
and our suggestion being correct, one of them or neither.
The inter-rater agreement between workers was 64\%.
Table~\ref{tbl:e2_eval} shows how these false positives were assessed.

\begin{table}[h]
  \centering\small
  \begin{tabular}{l|r}
  \toprule
    both & 4,506 (38.2\%) \\
    ground truth only & 3,768 (31.9\%)\\
    our suggestion only & 2,287 (19.4\%)\\
    neither & 1,242 (10.5\%)\\
    \midrule
    all & 11,803 (100\%)\\
    \bottomrule
  \end{tabular}
  \caption{\small{Relevant citation distribution for  \texttt{E2}.}}
  \label{tbl:e2_eval}
\end{table}

We see that in many cases our suggestion was equal to (38.2\%) or even
preferred (19.4\%) over the ground-truth suggestion. Hence, our method can even improve citation quality in Wikipedia. 

In the \texttt{E2} column in Table~\ref{tbl:t2_results_e1} we show the
updated results for $FC$ after collecting judgments for false positive
news articles. We see that for most of the types we have an average
gain of 18\% in terms of precision. We achieve the biggest gain of
28\% for the entity type \texttt{location}. For the types
\texttt{football player, creator, entertainer}, we can suggest news
citations with 90-91\% precision. Please note that we do not report
the recall score for \texttt{E2}, since assessing the appropriateness of 
every article in $\N$ as a citation for $s$ is not
feasible. The recall score is only reported w.r.t the ground-truth
articles in $\N^W_{13-15}$.

\section{Pipeline Evaluation}\label{sec:evaluation_pipe}

For the evaluation of both tasks in a pipeline scenario, we randomly
sample 1000 statements from all categories and ran the process of
citation discovery through both steps. Each statement is associated
with multiple entity types, as they are extracted from $e$ where
$T(e)$ is a set of types. For the \emph{statement categorization} task
we perform the evaluation based on our ground-truth; for the
\emph{citation discovery} we evaluate the suggested citations as in
evaluation strategy \texttt{E2}. Note, that here in the evaluation
pair we have a news article (that we suggest) and a resource that can
be of any type including \emph{book, web, journal}.

\textbf{Statement Categorization.}  We set up \emph{statement
  categorization} as a \emph{majority voting} categorization. For each
statement and the type specific classifiers $SC$ we predict the
category and pick the category that has the majority of votes. In
contrast to the \emph{statement categorization} in
Section~\ref{sec:statement_classification}, where the original task
aimed at showing for which types this task can be performed
accurately, we now aim to set up citation discovery in an automated
manner.

Based on the ground-truth, 340 out of the 1000 statements were
\emph{news statements}. We categorize 368 as news statements, out of
which 263 are correct, i.e. P=0.72 and R=0.77. It is interesting to
see that we can leverage additional information through \emph{majority
  voting}, where for the same statement and its associated types we
can predict with high accuracy the citation category label of $s$.

\textbf{Citation Discovery.} For the \emph{citation discovery} task we
ran it based on the generic $FC$ model trained on statements belonging
to all types, namely \texttt{owl:Thing}. We could use the type
specific $FC$, with additional costs for computing type specific
features.

In the second task, from the 368 statements classified as news
statements, we ran the citation discovery model $FC$. We are able to
suggest 549 news citations for 78 statements. Based on crowd-sourcing
evaluation, we suggest 346 relevant citations, i.e.  a precision of
P=0.63, out of which 200 citations are citations that were preferred
over existing ones in the ground-truth.  For 146 cases the citations
we suggest are considered to be equally appropriate as the existing
ones in the ground-truth, for 116 citations the ground-truth ones were
preferred over the ones we suggested. Note that our $FC$ models
suggest citations for $s$ only in case they fulfill the criteria in
Section~\ref{sec:missing_citations}, thus, enforcing high accuracy.

\section{Related Work}\label{sec:relatedwork}

\textbf{Citation Sources.} Ford et
al.~\cite{DBLP:conf/wikis/FordSMM13} analyze the citation behavior of
Wikipedia editors with respect to their adherence to the citation
guidelines. They investigate what types of sources are most often
cited, i.e. \emph{primary, secondary} and \emph{tertiary} as defined
in
Wikipedia\footnoteref{reliability}. They
conclude that news are one of the top cited source in the
\emph{secondary} type, while they see a growing trend of
\emph{primary} sources due to their persistence on the web, contrary
to the policies of preferring secondary sources. Luyt and
Tan~\cite{DBLP:journals/jasis/LuytT10} analyze a subset of history
entity pages and show that citations are biased towards a specific
group of sources.
\cite{DBLP:conf/wikis/FordSMM13,DBLP:journals/jasis/LuytT10} emphasize
the importance of citations in Wikipedia as a means to ensure the
quality of entity pages.

\textbf{Wikipedia Quality.} Anderka et
al.~\cite{DBLP:conf/sigir/AnderkaSL12} propose an approach to predict
quality flaws in Wikipedia pages. A quality flaw in Wikipedia is
usually annotated with specific \emph{cleanup} tags. They train a
model to predict quality flaws, where among the top--10 quality flaws
they identify \emph{unreferenced, refimprove, primary sources} as some
of the most serious flaws. Our work is complementary to theirs since
we aim at finding appropriate citations for Wikipedia statements,
thereby improving the quality of Wikipedia pages.

\textbf{Wikipedia Enrichment.} Sauper and Barzilay
\cite{DBLP:conf/acl/SauperB09} propose an approach to automatically
generate complete entity pages for a specific entity type. The
approach is trained on already-populated entity pages of a specific
type by learning templates about the section structure at the type
level. For a new entity page, they extract documents through Web
search (with entity and section title as a query) and identify the
most relevant paragraphs to add in a section. Fetahu et al. in~\cite{DBLP:conf/cikm/FetahuMA15} proposed an approach for
suggesting news articles for a Wikipedia entity and entity section. They
first identify news articles that are important to an entity and in
which the entity is salient, and further identify the most appropriate
section to suggest the article. In case of a missing section, a new
section is added by exploiting the section structure from the entity
type.

This work differs from \cite{DBLP:conf/acl/SauperB09,DBLP:conf/cikm/FetahuMA15} as we do not add content or suggest news articles to a complete section in an
entity page, but rather provide citations to already existing
statements.

\textbf{Cumulative Citation Recommendation (CCR).} TREC introduced the CCR track in the Knowledge base acceleration track in 2012. For a stream of news and social media content and a target entity from a knowledge base (Wikipedia), the goal of the task is to generate a score for each document based on how pertinent it is to the input entity. Balog et
al.~\cite{DBLP:conf/riao/BalogRTN13,DBLP:conf/sigir/BalogR13} propose
approaches that find entity mentions in the document collection and
rank them according to how central the entity is in the respective
documents. This however is a filtering task for documents towards checking if they are relevant for a pre-defined set of entities. In contrast, in our task we aim at finding news citations
as evidence for Wikipedia statements.

\section{Conclusions}
\label{sec:conclusion}

In this work we define and attempt to solve the automatic news
citation discovery problem for Wikipedia. We define two tasks --
\emph{sentence categorization} and the \emph{citation discovery} --
towards finding the correct news citation for a given Wikipedia
statement. For the sentence categorization task, we learn a
multi-class classifier to predict if a statement requires a news
statement. For the news citation discovery problem, we first find the
likely candidates by a retrieval model over a real-world news
collection followed by a binary classification for the top-ranked
candidates.

We find that statement categorization is a hard problem
due to lack of context for the NLP-based features to perform
well. However, the Wikipedia page and its type structure provide
important cues towards accurate classification. On the other hand, we
perform well on the citation discovery task with 67\% precision (for
top-categories) using the automated evaluation, which further improves to
over 80\% when crowd-sourced. This shows that we not only identify the
correct ground truth articles present in Wikipedia, but in some cases
our suggestions are a better fit compared to the sources in Wikipedia.

\section*{Acknowledgments.} This work is funded by the ERC Advanced Grant ALEXANDRIA (grant no. 339233).

\end{document}